\documentclass[10pt,a4papper]{article}
\usepackage{indentfirst}
\usepackage[]{graphicx}
\usepackage{geometry} 
\begin{document}

\title{\textbf{The decay $\tau \rightarrow K^{0}K^{-}\nu_{\tau}$ in the extended Nambu-Jona-Lasinio model}}
\author{M. K. Volkov\footnote{volkov@theor.jinr.ru}, A. A. Pivovarov\footnote{tex$\_$k@mail.ru}\\
\small
\emph{Bogoliubov Laboratory of Theoretical Physics, JINR, Dubna, 141980, Russia}}
\maketitle
\small

\begin{abstract}
The full and differential widths of the decay $\tau \rightarrow K^{0}K^{-}\nu_{\tau}$ are calculated in the framework of the extended Nambu-Jona-Lasinio model.
The contributions of the subprocesses with the intermediate vector mesons $\rho$(770) and $\rho$(1450) are taken into account. The obtained results are in
satisfactory agreement with the experimental data.
\end{abstract}
\large
\section{Introduction}
The present paper finishes the series of works devoted to describing of $\tau$-decays in neutrino and two pseudoscalar mesons
in the framework of the extended Nambu-Jona-Lasinio (NJL) model
\cite{Volkov:1996br, Volkov:1996fk, Volkov:1997dd, Volkov:1999yi, Volkov:2006vq}. Indeed, recently, the widths of the decays
$\tau \rightarrow (\pi, \pi(1300)) \nu_{\tau}$ \cite{Ahmadov:2015zua}, $\tau \rightarrow (\eta, \eta') \pi \nu_{\tau}$
\cite{Volkov:2012be}, $\tau \rightarrow K^{-} \pi^{0} \nu_{\tau}$ \cite{Volkov:2015vij},
$\tau \rightarrow (\eta, \eta') K^{-} \nu_{\tau}$ \cite{Volkov:2016gsi} were calculated by using this model without applying any additional
arbitrary parameters. This approach differs the NJL model \cite{Volkov:2006vq, Volkov:1986zb, Ebert:1985kz, Ebert:1994mf} from some other phenomenological models used for describing such processes
\cite{Finkemeier:1996dh, Li:1996md, Palomar:2002hp, Nussinov:2009sn, Czyz:2010hj, Dubnicka:2010te, Paver:2011md, Dumm:2012vb, Escribano:2014joa, Escribano:2016ntp}.
As a rule, these models use the vector dominance method, chiral symmetry and series of arbitrary parameters which are different
for various processes.

\section{The Lagrangian of the extended NJL model for the mesons \\
$K^{0,\pm}, \rho^{\pm}$ and their first radially excited states}
In the extended NJL model, the quark-meson interaction Lagrangian for pseudoscalar $K^{0,\pm}$,
vector $\rho^{\pm}$ mesons and their first radially excited states takes the form:

\begin{equation}
\Delta L_{int}(q,\bar{q},K,\rho) = \bar{q}\left[i\gamma^{5}\sum_{j =0, \pm}\lambda_{j}^{K}(a_{K}K^{j} + b_{K}K^{'j})
+ \frac{1}{2}\gamma^{\mu}\sum_{j = \pm}\lambda_{j}^{\rho}(a_{\rho}\rho^{j}_{\mu} + b_{\rho}\rho^{'j}_{\mu})\right]q,
\end{equation}
where $q$ and $\bar{q}$ are the u-, d- and s- constituent quark fields with masses $m_{u} = m_{d} = 280$MeV,
$m_{s} = 420$MeV \cite{Volkov:1999yi},\cite{Volkov:2001ns}, $K^{0,\pm}$, $\rho^{\pm}$ are
the pseudoscalar and vector mesons, the excited states are marked with prime,

\begin{displaymath}
a_{a} = \frac{1}{\sin(2\theta_{a}^{0})}\left[g_{a}\sin(\theta_{a} + \theta_{a}^{0}) +
g_{a}^{'}f_{a}(\vec{k}^{2})\sin(\theta_{a} - \theta_{a}^{0})\right],
\end{displaymath}
\begin{equation}
\label{Coefficients}
b_{a} = \frac{-1}{\sin(2\theta_{a}^{0})}\left[g_{a}\cos(\theta_{a} + \theta_{a}^{0}) +
g_{a}^{'}f_{a}(\vec{k}^{2})\cos(\theta_{a} - \theta_{a}^{0})\right],
\end{equation}
$f\left(\vec{k}^{2}\right) = 1 + d \vec{k}^{2}$ is the form factor for description of the first radially excited states
\cite{Volkov:1996br},\cite{Volkov:1996fk}, $d$ is the slope parameter, $\theta_{a}$ and $\theta_{a}^{0}$ are
the mixing angles for the strange mesons in the ground and excited states

\begin{displaymath}
d_{uu} = -1.784 \textrm{GeV}^{-2}, \quad d_{us} = -1.761 \textrm{GeV}^{-2},
\end{displaymath}
\begin{equation}
\begin{array}{cc}
\theta_{K} = 58.11^{\circ},      & \theta_{\rho} = 81.8^{\circ},\\
\theta_{K}^{0} = 55.52^{\circ},  & \theta_{\rho}^{0} = 61.5^{\circ}.
\end{array}
\end{equation}

The matrices

\begin{displaymath}
\lambda_{+}^{K} = \sqrt{2} \left(\begin{array}{ccc}
0 & 0 & 1\\
0 & 0 & 0\\
0 & 0 & 0
\end{array}\right), \quad
\lambda_{-}^{K} = \sqrt{2} \left(\begin{array}{ccc}
0 & 0 & 0\\
0 & 0 & 0\\
1 & 0 & 0
\end{array}\right), \quad
\lambda_{0}^{K} = \sqrt{2} \left(\begin{array}{ccc}
0 & 0 & 0\\
0 & 0 & 0\\
0 & 1 & 0
\end{array}\right),
\end{displaymath}

\begin{equation}
\lambda_{+}^{\rho} = \sqrt{2} \left(\begin{array}{ccc}
0 & 1 & 0\\
0 & 0 & 0\\
0 & 0 & 0
\end{array}\right), \quad
\lambda_{-}^{\rho} = \sqrt{2} \left(\begin{array}{ccc}
0 & 0 & 0\\
1 & 0 & 0\\
0 & 0 & 0
\end{array}\right).
\end{equation}

The coupling constants:

\begin{displaymath}
g_{K} = \left(\frac{4}{Z_{K}}I_{2}(m_{u},m_{s})\right)^{-1/2} \approx 3.77,
\quad g_{K}^{'} = \left(4I_{2}^{f_{us}^{2}}(m_{u},m_{s})\right)^{-1/2} \approx 4.69,
\end{displaymath}
\begin{equation}
\label{Constants}
g_{\rho} = \left(\frac{2}{3}I_{2}(m_{u},m_{u})\right)^{-1/2} \approx 6.14,
\quad g_{\rho}^{'} = \left(\frac{2}{3}I_{2}^{f_{uu}^{2}}(m_{u},m_{u})\right)^{-1/2} \approx 9.87,
\end{equation}
where

\begin{equation}
Z_{K} = \left(1 - \frac{3}{2}\frac{(m_{u} + m_{s})^{2}}{M^{2}_{K_{1}}}\right)^{-1} \approx 1.83,
\end{equation}
$Z_{K}$ is the factor corresponding to the $K - K_{1}$ transitions,
$M_{K_{1}} = 1272$MeV \cite{Agashe:2014kda} is the mass
of the axial-vector $K_{1}$ meson, and the integral $I_{2}$ has the following form:

\begin{equation}
I_{2}^{f^{n}}(m_{1}, m_{2}) =
-i\frac{N_{c}}{(2\pi)^{4}}\int\frac{f^{n}(\vec{k}^{2})}{(m_{1}^{2} - k^2)(m_{2}^{2} - k^2)}\theta(\Lambda_{3}^{2} - \vec{k}^2)
\mathrm{d}^{4}k,
\end{equation}
$\Lambda_{3} = 1.03$ GeV is the cut-off parameter \cite{Volkov:1999yi}.

The all these parameters were calculated earlier and are standard for the extended NJL model \cite{Volkov:1996fk, Volkov:1999yi}.

\section{The amplitude of the decay $\tau \rightarrow K^{0}K^{-}\nu_{\tau}$ in the extended NJL model}
\begin{figure}[h]
\center{\includegraphics[scale = 0.7]{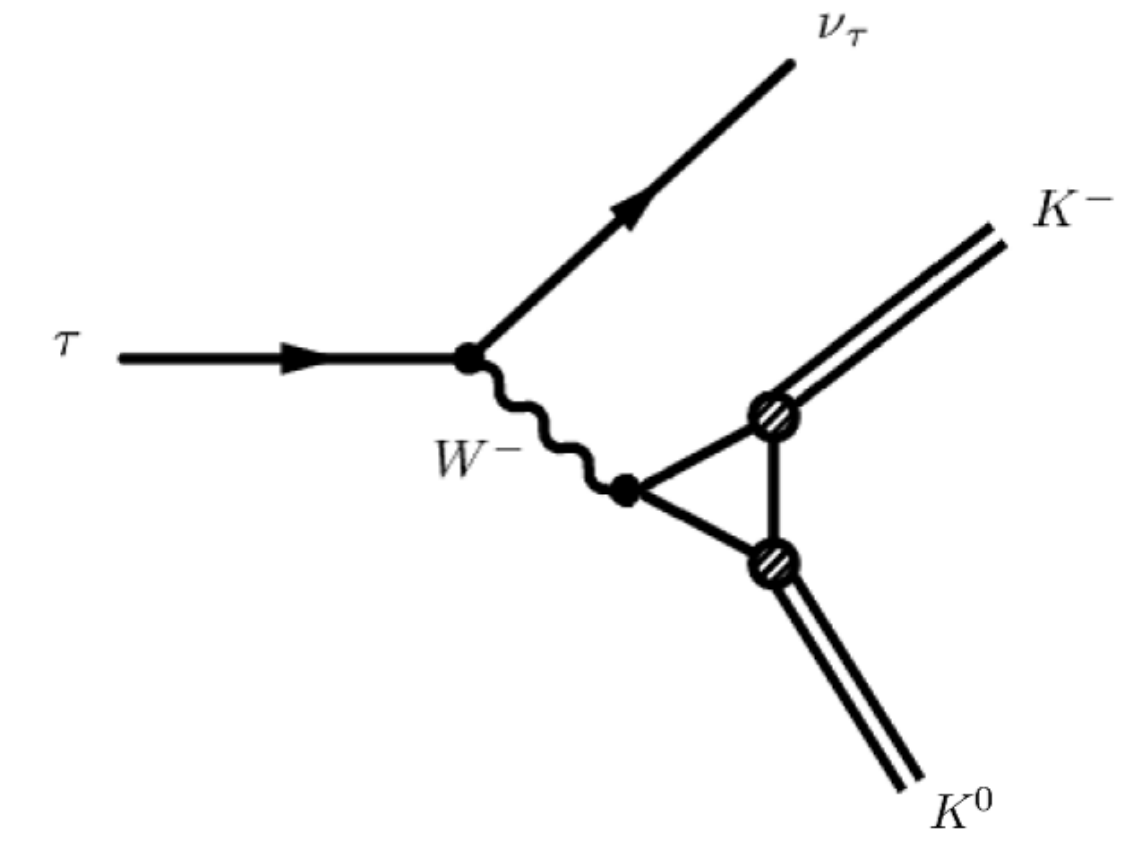}}
\caption{The decay $\tau \rightarrow K^{0}K^{-}\nu_{\tau}$ with the intermediate $W$-boson (Contact diagram)}
\label{Contact}
\end{figure}

\begin{figure}[h]
\center{\includegraphics[scale = 0.9]{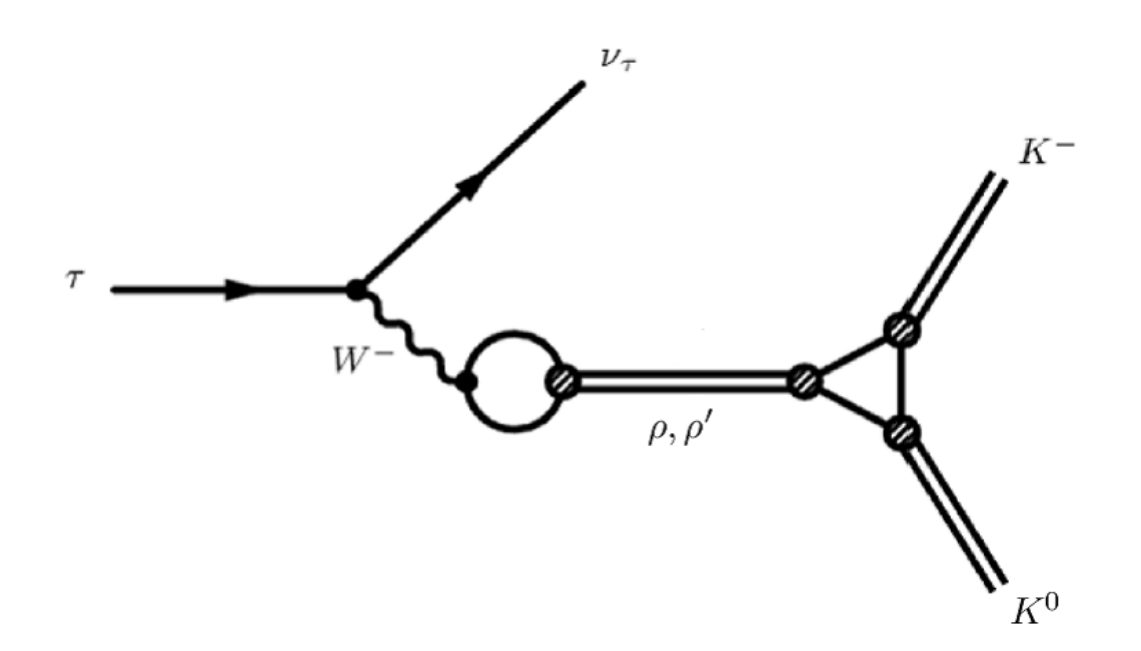}}
\caption{The decay $\tau \rightarrow K^{0}K^{-}\nu_{\tau}$ with the intermediate vector $\rho(770)$ and $\rho(1450)$ mesons}
\label{Intermediate}
\end{figure}

The diagrams of the process $\tau \rightarrow K^{0}K^{-}\nu_{\tau}$ are shown in Figs.\ref{Contact},\ref{Intermediate}.


The amplitude of this process takes the form:

\begin{displaymath}
T = -2\sqrt{2}i G_{F}|V_{ud}|l^{\mu}
\left\{I_{KK}g_{\mu\nu} + \frac{I_{KK\rho}C_{\rho}}{g_{\rho}}\cdot\frac{g_{\mu\nu}q^{2} - q_{\mu}q_{\nu}}
{M_{\rho}^{2} - q^{2} - i\sqrt{q^{2}}\Gamma_{\rho}}\right.
\end{displaymath}
\begin{equation}
\left. + \frac{I_{KK\rho^{'}}C_{\rho^{'}}}{g_{\rho}}\cdot\frac{g_{\mu\nu}q^{2} - q_{\mu}q_{\nu}}
{M_{\rho^{'}}^{2} - q^{2} - i\sqrt{q^{2}}\Gamma_{\rho^{'}}}\right\} (p_{K^{0}} - p_{K^{-}})^{\nu},
\end{equation}
where $G_{F} = 1.16637 \cdot 10^{-11}$MeV$^{-2}$ is the Fermi constant, $V_{ud} = 0.97428$ is the element of the Cabbibo-Kobayashi-Maskawa matrix,
$l^{\mu} = \bar{\nu}_{\tau}\gamma^{\mu}\tau$ is the lepton current, $q = p_{K^{0}} - p_{K^{-}}$, $M_{\rho} = 775.11 \pm 0.34$MeV,
$M_{\rho^{'}} = 1465 \pm 25$MeV, $\Gamma_{\rho} = 149.1 \pm 0.8$MeV, $\Gamma_{\rho^{'}} = 400 \pm 60$MeV
are the masses and the full widths of the vector mesons \cite{Agashe:2014kda}.

The first term corresponds to the diagram with the intermediate $W$-boson, the second and third terms correspond
to the diagrams with the intermediate vector mesons $\rho(770)$ and $\rho(1450)$. The numerical coefficients

\begin{displaymath}
C_{\rho} = \frac{1}{\sin\left(2\theta_{\rho}^{0}\right)}\left[\sin\left(\theta_{\rho} + \theta_{\rho}^{0}\right) +
R_{V}\sin\left(\theta_{\rho} - \theta_{\rho}^{0}\right)\right],
\end{displaymath}
\begin{displaymath}
C_{\rho'} = \frac{-1}{\sin\left(2\theta_{\rho}^{0}\right)}\left[\cos\left(\theta_{\rho} + \theta_{\rho}^{0}\right) +
R_{V}\cos\left(\theta_{\rho} - \theta_{\rho}^{0}\right)\right],
\end{displaymath}
\begin{displaymath}
R_{V} = \frac{I_{2}^{f}(m_{u},m_{u})}{\sqrt{I_{2}(m_{u},m_{u})I_{2}^{f^{2}}(m_{u},m_{u})}}.
\end{displaymath}
\begin{equation}
I_{abc} =
-i\frac{N_{c}}{(2\pi)^{4}}\int\frac{a(\vec{k}^{2})b(\vec{k}^{2})c(\vec{k}^{2})}{(m_{s}^{2} - k^2)(m_{u}^{2} - k^2)}
\theta(\Lambda_{3}^{2} - \vec{k}^2) \mathrm{d}^{4}k,
\end{equation}
where $a(\vec{k}^{2})$, $b(\vec{k}^{2})$ and $c(\vec{k}^{2})$ are the coefficients from the Lagrangian defined in (\ref{Coefficients}).

\section{Numerical estimations}
The calculated branching of the process $\tau \rightarrow K^{0}K^{-}\nu_{\tau}$ is
\begin{equation}
Br(\tau \rightarrow K^{0}K^{-}\nu_{\tau}) = 12.7 \cdot 10^{-4}.
\end{equation}

The experimental value of this branching are
\begin{equation}
Br(\tau \rightarrow K^{0}K^{-}\nu_{\tau})_{exp} = (14.9 \pm 0.5) \cdot 10^{-4} \textrm{ \cite{Agashe:2014kda}}
\end{equation}
\begin{equation}
Br(\tau \rightarrow K^{0}K^{-}\nu_{\tau})_{exp} = (15.1 \pm 4.3) \cdot 10^{-4} \textrm{ \cite{Coan:1996iu}}.
\end{equation}

The mass and full decay width of the meson $\rho$(1450) are not defined precisely.
It is interesting to note that if we choose the minimal values of mass and full width of
this meson ($M_{\rho^{'}} = 1440$MeV, $\Gamma_{\rho^{'}} = 340$MeV), the results are in better agreement with the experimental data:
\begin{equation}
Br(\tau \rightarrow K^{0}K^{-}\nu_{\tau}) = 14.7 \cdot 10^{-4}.
\end{equation}

\begin{figure}[h]
\center{\includegraphics[scale = 0.9]{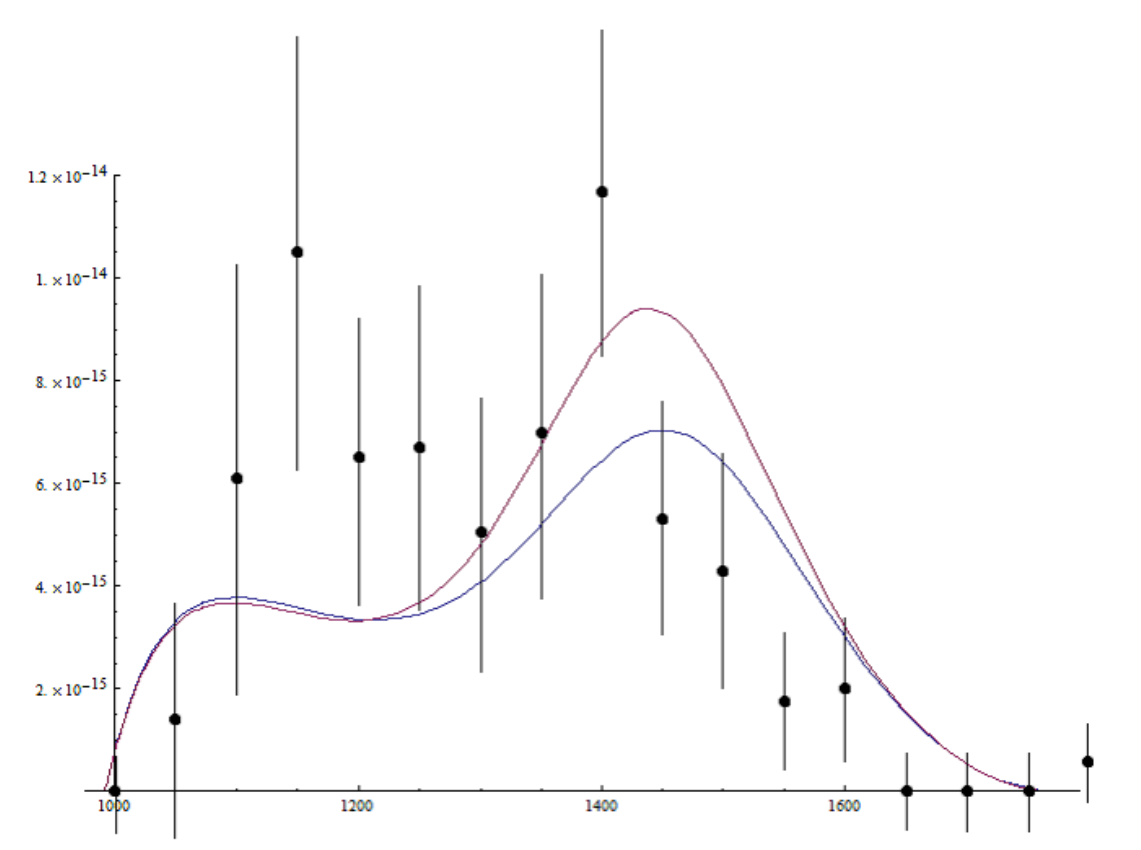}}
\caption{Differential width of the decay $\tau \rightarrow K^{0}K^{-}\nu_{\tau}$}
\label{Diff}
\end{figure}

The comparison of the calculated and experimental differential width is shown in Fig.~\ref{Diff}.
The solid lines correspond to our theoretical differential width. The blue one is for the case of middle values of
mass and full width of the meson $\rho$(1450), the red one is for the case of minimal values of them.
The points correspond to the experimental values \cite{Coan:1996iu}.

\section{Conclusion}

From two options considered in this paper, it is easy to see that the results are sensitive to the choice
of the full decay width and mass of $\rho$(1450) meson. Since these mass and width are defined with the large errors, we can choose
the minimal allowable values. This leads to better agreement with the experimental data. It is interesting to compare our results
with the results from other phenomenological models and experiments. Such comparison is shown in Tab.~\ref{TabBr}.
Our results are in satisfactory agreement with the experimental data.

\begin{table}[h]
\caption{The branching ratios for the process $\tau \rightarrow K^{0}K^{-}\nu_{\tau}$.}
\label{TabBr}
\begin{center}
\begin{tabular}{|c|c|c|}
\hline
            & Br $(\times 10^{-4})$ & References                                          \\
\hline
Theory      &            27         & B.A. Li \cite{Li:1996md}                            \\
            &     12.5 $\pm$ 1.3    & J.E. Palomar \cite{Palomar:2002hp}                  \\
            &        13.5/19        & H.Czyz, A. Grzelinska, J.H. Kuhn \cite{Czyz:2010hj} \\
            &           16          & S. Dubnicka, A.Z. Dubnickova \cite{Dubnicka:2010te} \\
            &        12.7(14.7)     & Our result                                          \\
\hline
Experiment  &    15.1 $\pm$ 4.3     & CLEO \cite{Coan:1996iu}                             \\
            &    16.2 $\pm$ 3.2     & ALEPH \cite{Barate:1999hi}                          \\
            &    14.8 $\pm$ 0.68    & Belle \cite{Ryu:2014vpc}                            \\
            &    14.9 $\pm$ 0.5     & PDG \cite{Agashe:2014kda}                           \\
\hline
\end{tabular}
\end{center}
\end{table}

\section*{Acknowledgments}
We are grateful to A. B. Arbuzov and O. V. Teryaev for useful discussions.

\end{document}